\documentclass[12pt]{iopart}

\usepackage{graphicx}

\begin{document}

\title{Improved Superconducting Properties in Bulk MgB$_{2}$ Prepared by High Energy Milling of Mg and B Powder}

\author{Y.~F.~Wu$^1$\footnote[1]{To whom correspondence should be addressed
(wyf7777@tom.com)}, Y.~F.~Lu$^1$, G.~Yan$^1$, J.~S.~Li$^2$,
Y.~Feng$^1$, H.~P.~Tang$^1$, S.~K.~Chen$^1$, H.~L.~Xu$^3$,
C.~S.~Li$^1$ and P.X.Zhang$^1$}

\address{$^1$ Northwest Institute for Nonferrous Metal Research, P. O. Box 51, Xian, Shaanxi 710016, P. R. China}
\address{$^2$ Northwestern Polytechnical University, Xi'an 710012, P.R.China}
\address{$^3$ School of Material Science and Engineering, Zhengzhou University, Zhengzhou, Henan 450002, People¡¯s Republic of
China}

\begin{abstract}
The MgB$_{2}$ bulks were prepared by high energy milling of Mg and B
powder. The correlations among milling times, microstructure and
superconducting properties were investigated in MgB$_{2}$ bulks.
Samples were characterized by x-ray diffraction (XRD), energy
dispersive spectrometry (EDX) and scanning electron microscope
(SEM), and the magnetization properties were examined by a
Superconducting quantum interfere device (SQUID) magnetometer. It
showed that the high energy milling is an effective approach to get
fine crystalline (40-100nm) bulk MgB$_{2}$ with good grain
connectivity and high \textit{J$_{c}$} performance. The critical
current density reaches to 2.0$\times\,$10$^{6}$A/cm$^{2}$ at 15K
and 0.59T, 5.7¡Á$\times\,$10$^{5}$A/cm$^{2}$ at 2T and
3.0¡Á$\times\,$10$^{4}$A/cm$^{2}$ at 5T.

\end{abstract}



















\maketitle

\section{Introduction}

The improvement of the intrinsic properties of MgB$_{2}$ was
recognized as a decisive goal to enable potential applications[1].
Conventional powders and sintered polycrystalline bulk MgB$_{2}$
samples typically exhibit deteriorated superconducting properties
due to weak pinning[2-3]. The mechanically alloyed (MA) samples have
about 1000 times smaller grains than hot deformed samples and
samples sintered at high pressure[4]. The observed increased
\textit{H$_{irr}$(T)} and high \textit{J$_{c}$(H)} manifest improved
flux pinning of MA samples, which attribute to small grains and the
enhanced number of grain boundaries for the nanocrystalline
material[4]. It has been demonstrated before that the
nanocrystalline sample has distinctly higher irreversibility fields
than the bulk sample with micrometer-sized grains, especially at the
low temperature[5]. One possible reason for the strong pinning found
for thin films as well is their small grain size of about 10nm[5].
The mechanical alloying (MA) technique for MgB$_{2}$ powder
preparation is expected for obtaining enhanced magnetic flux pinning
by microstructure refinement. However, it costs as long as
20~100h[4,6,7 -12] for in situ MA precursor powder preparation. The
use of short-time unalloyed high energy milling of Mg and B powder
as precursor material represents an efficient combination between
conventional powder preparation and mechanical alloying techniques.
Agate milling media was chosen for dispelling the bad effect of the
impurities. Additionally, MgB$_{2}$ is a line compound[13] and a
certain deviation of the stoichiometry lead to substantially
increased critical current densities compared to
the stoichiometric composition[11]. The precursor powder with a Mg excess of 5%
was employed according to the literature[11,14-16].

\section{Experimental details}

Mg (99.8\%) and amorphous B (95\%) powder with 5
were filled under purified Ar-atmosphere into an agate milling
container and milling media. The milling was performed on a SPEX
8000M mill for different times t$_{m}$ =1, 5, 10h using a
ball-to-powder mass ratio of 3. The milled powders were then cold
pressed to form pellets with a diameter of 20mm and a height of 3mm.
The pellets were placed in an alumina crucible inside a tube furnace
under ultra-high purity Ar-atmosphere. The heat treatment parameters
were optimally chosen at 750¡æ and 1h from our previous work[17],
then cooled down to the room temperature.

The phase compositions of the samples were characterized by the
APD1700 X-ray diffraction instrument. The surface morphology and
microstructures of the samples were characterized by the JSM-6460
and the JSM-6700F scanning electron microscope.

A SQUID magnetometer by Quantum Design was used to measure the AC
magnetic susceptibility of the samples over a temperature range of 5
to 50 K under an applied field of 1Oe. Magnetization versus magnetic
field (\textit{M-H}) curves were also measured on rectangular-shaped
samples at temperatures of 10K and 15K under magnetic fields up to
90000Oe to determine the critical current density
\textit{J$_{c}$(H)}.

\section{Results and discussion}

XRD and EDX analysis reveal the appearance of a small amount of MgO
impurity for both as-milled and sintered samples. The relative
percentage composition of MgO phase is gradually increased with
prolonged milling time for both samples. It is most probably because
of the oxide diffusing into the grain surfaces during the particle
reduction process and unavoidable oxygen traces during the sintering
process.

The microstructures of the investigated samples are shown in Fig.1.
Scanning electron microscope images mainly show spherical grains of
about 40~100 nm in size for 5h milled sample. The impurity phases
are evidently observed for sample milling for 10h, as marked by
black arrows. Fig.2 indicates the distribution of the second phases
for differently milled bulk MgB$_{2}$. There are a few small dark
grey impurity phases for the sample milling for 1h. The sample
milling for 5h has few impurity phases. However, there are a large
number of stripped impurity phases for the sample milling for 10h.

The magnetic field dependence of magnetization for samples milling
for different times is shown in Fig.3. \textit{J$_{c}$(H)} was
calculated by Bean critical state model from magnetization curves
(see Fig.4). Shown in inset is the irreversible field as a function
of milling time. \textit{H$_{irr}$} values were determined from the
closure of hyseresis loops with a criterion of
\textit{J$_{c}$}=10$^{2}$A/cm$^{2}$[18]. As we can see, the sample
milling for 5h has a significantly higher \textit{H$_{irr}$} and
\textit{J$_{c}$} than the other samples in magnetic field. The
critical current density reach to 2.0$\times\,$10$^{6}$A/cm$^{2}$ at
15K and 0.59T, 5.7$\times\,$10$^{5}$A/cm$^{2}$ at 2T and
3.0$\times\,$10$^{4}$A/cm$^{2}$ at 5T. The improved pinning of this
material seems to be caused by enhanced grain boundary pinning
provided by the large number of grain boundaries in the sample.

The maximum \textit{J$_{c}$} was not firstly discovered in MgB$_{2}$
from ball milled precursor powder. Flukiger's group[19] declaimed
that \textit{J$_{c}$} of ex situ Fe/MgB$_{2}$ tapes with ball milled
powder first shows an enhancement for a particles size of 3/30¦Ìm,
followed by a decrease for further reductions to 1.5/10¦Ìm and 1¦Ìm.
They thought that the maximum of \textit{J$_{c}$} is a compromise
between enhanced flux pinning at the grain boundaries, caused by
these chemical impurities at a nanoscale and a decrease of
\textit{J$_{c}$}, caused by the introduction of too many impurities,
reducing percolation of the current. However, O. Perner's group[10]
considered there are two competing processes taking place during the
high-energy milling. The first one initiates an improvement of the
superconducting properties of bulk MgB$_{2}$. It attributes to the
grain refinement resulting in a higher reactivity and, therefore, an
optimal grain connectivity and high density of grain boundaries in
MgB$_{2}$ bulks. The second process is expected to be the
introduction of oxygen from the working atmosphere with increasing
milling time, causing increased content of impurity phases with a
reduced grain connectivity. The experimental data of ours are
evidently in support of O. Perner's opinion from IFW Dresden.

Fig.5 shows field dependence of the volume pinning force
\textit{F$_{p}$(H)} at 15K for samples heated at
$750\,^{\circ}\mathrm{c}$ and milled for different times.
\textit{F$_{p}$(H)} is normalized by the maximum volume pinning
force \textit{F$_{p.max }$} at the same temperature and for
different milling times. The shapes of these profiles depend on the
milling time. For 5h milled sample, the
\textit{F$_{p}$(H)}/\textit{F$_{p.max}$} values were obviously
larger than those of other samples over 1T, indicating enhanced flux
pinning in high field region. The \textit{F$_{p.max }$} values of 1h
and 5h milled samples reach 11.3 and 13.4GNm$^{-3}$ at 15K. These
values are nearly in the range of those commercial superconductors,
NbTi and Nb$_{3}$Sn, which show 15~30GNm$^{-3}$ at 4.2K[20].

Figure 6 shows a comparison of magnetic \textit{J$_{c}$(H)} for a 5h
milled sample with data reported in literature[21-25].
\textit{J$_{c}$} for this sample exhibits a better field performance
and higher values of \textit{J$_{c}$}. In the magnetic field lower
than 3.5T, our 5h milled sample shows the highest \textit{J$_{c}$}
in 15K. The best \textit{J$_{c}$} for the 5h milled sample was
2.3$\times\,$10$^{5}$A/cm$^{2}$ at 3T and 15K, which exceeds the
\textit{J$_{c}$} values of state-of-the-art Ag/Bi-2223 tapes. Our 5h
milled MgB$_{2}$ sample is even comparable with SiC doped MgB$_{2}$
bulks by Dou's group[23], which had exhibited the strongest reported
flux pinning and the highest J$_{c}$ in high fields to date.

In conclusion, we succeeded in preparing high critical current
density bulk MgB$_{2}$ by high energy ball-milling technique. It
demonstrated that it is an effective approach to get fine
crystalline bulk MgB$_{2}$ with good grain connectivity and
extraordinary high J$_{c}$ performance. The flux pinning is enhanced
in our sample by a large number of grain boundaries. Further
improvement is expected to be in doped samples using high energy
ball-milling technique.

\section{Acknowledgment}
This work was partially supported by National Natural Science
Foundation project (Contract No. 50472099) and National Basic
Research Program of China (Contract No. 2006CB601004).

\newpage
\textbf{Figure captions}

Fig.1 Fig.1 The SEM photograph of the MgB$_{2}$ bulks for different
milling times: (a) 1h   (b) 5h   (c) 10h.

Fig.2 The distribution of the second phase in bulk MgB2 for
different milling times: (a) 1h   (b) 5h   (c) 10h.

Fig.3 Magnetization \textit{M} as a function of magnetic field
\textit{H} at 15K for the samples heated at
$750\,^{\circ}\mathrm{c}$ and milled for different times.

Fig.4 Magnetization critical current density \textit{J$_{c}$} as a
function of magnetic field \textit{H} at 15 K for the samples heated
at $750\,^{\circ}\mathrm{c}$ and milled for different times. Shown
in inset is the irreversible field as a function of milling time.
\textit{H$_{irr}$} values were determined from the closure of
 hyseresis loops with a criterion of \textit{J$_{c}$}=10$^{2}$A/cm$^{2}$[18].

Fig.5 Field dependence of the volume pinning force
\textit{F$_{p}$(H)} at 15K for samples heated at
$750\,^{\circ}\mathrm{c}$ and milled for different times.
\textit{F$_{p}$(H)} is normalized by the maximum volume pinning
force \textit{F$_{p.max}$} at the same temperature and milling time.

Fig.6 A comparison of magnetic \textit{J$_{c}$(H)} for our 5h milled
sample and for samples that were: MA with C doping by Uni. Of WM(see
Ref. 21),MA with 5\% Mg Surplus by IFW(see Ref. 22), undoped (see
Ref. 23) and doped samples(see Ref. 23-25) by Dou's group.

\newpage
Fig.1
\begin{figure}[hp]
\centering
\includegraphics[width=180pt]{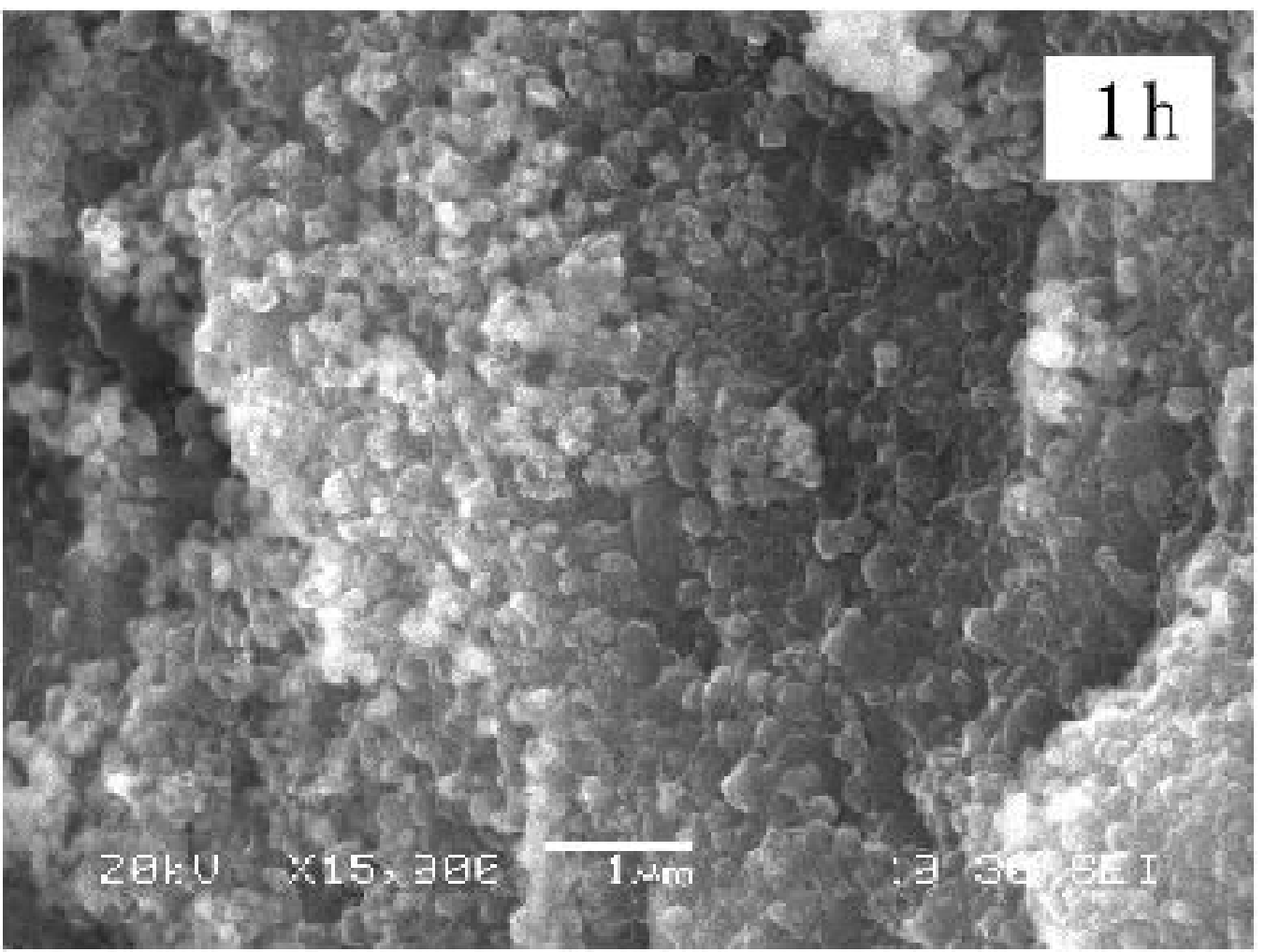}%
\includegraphics[width=180pt,height=135pt]{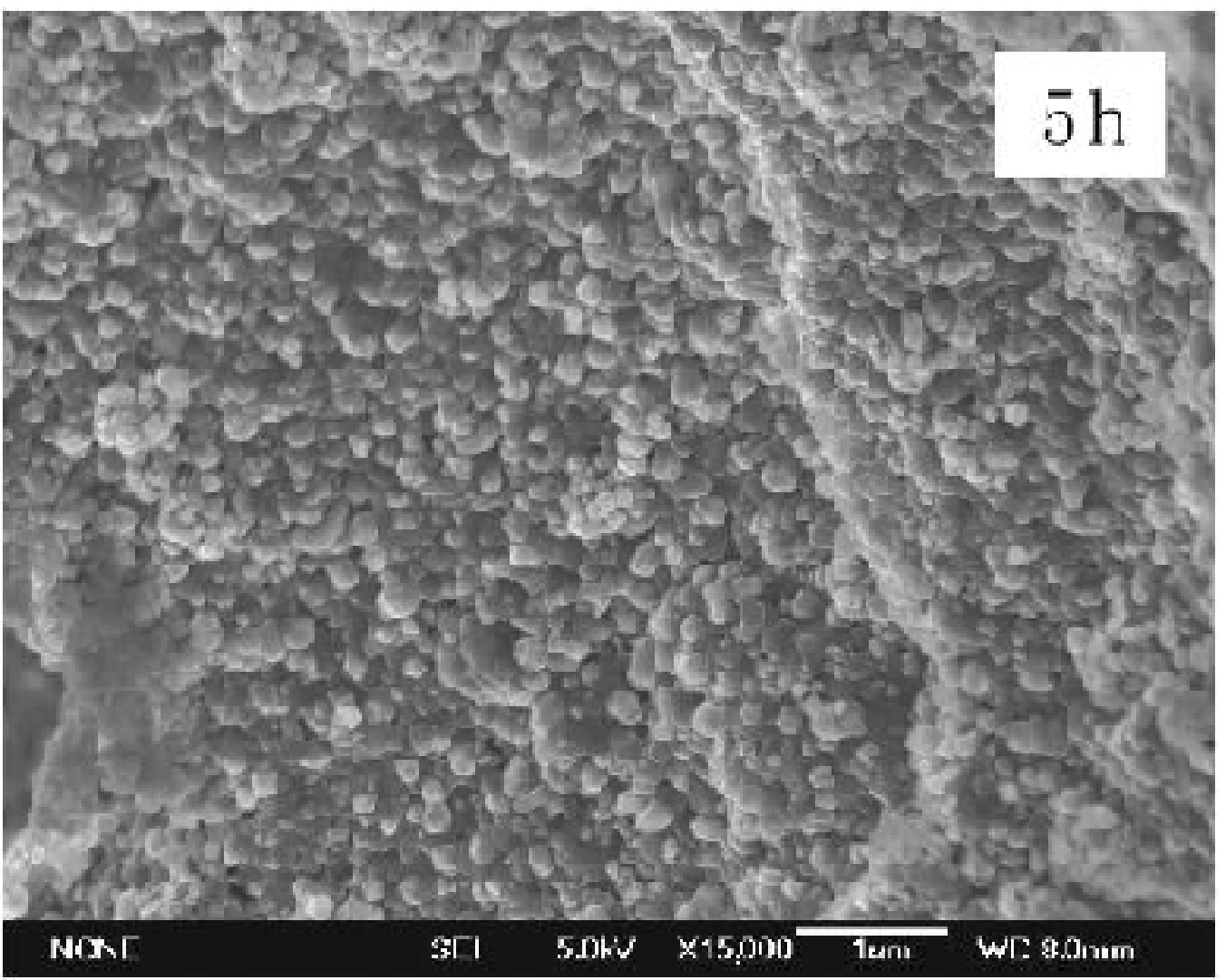}
\includegraphics[width=180pt]{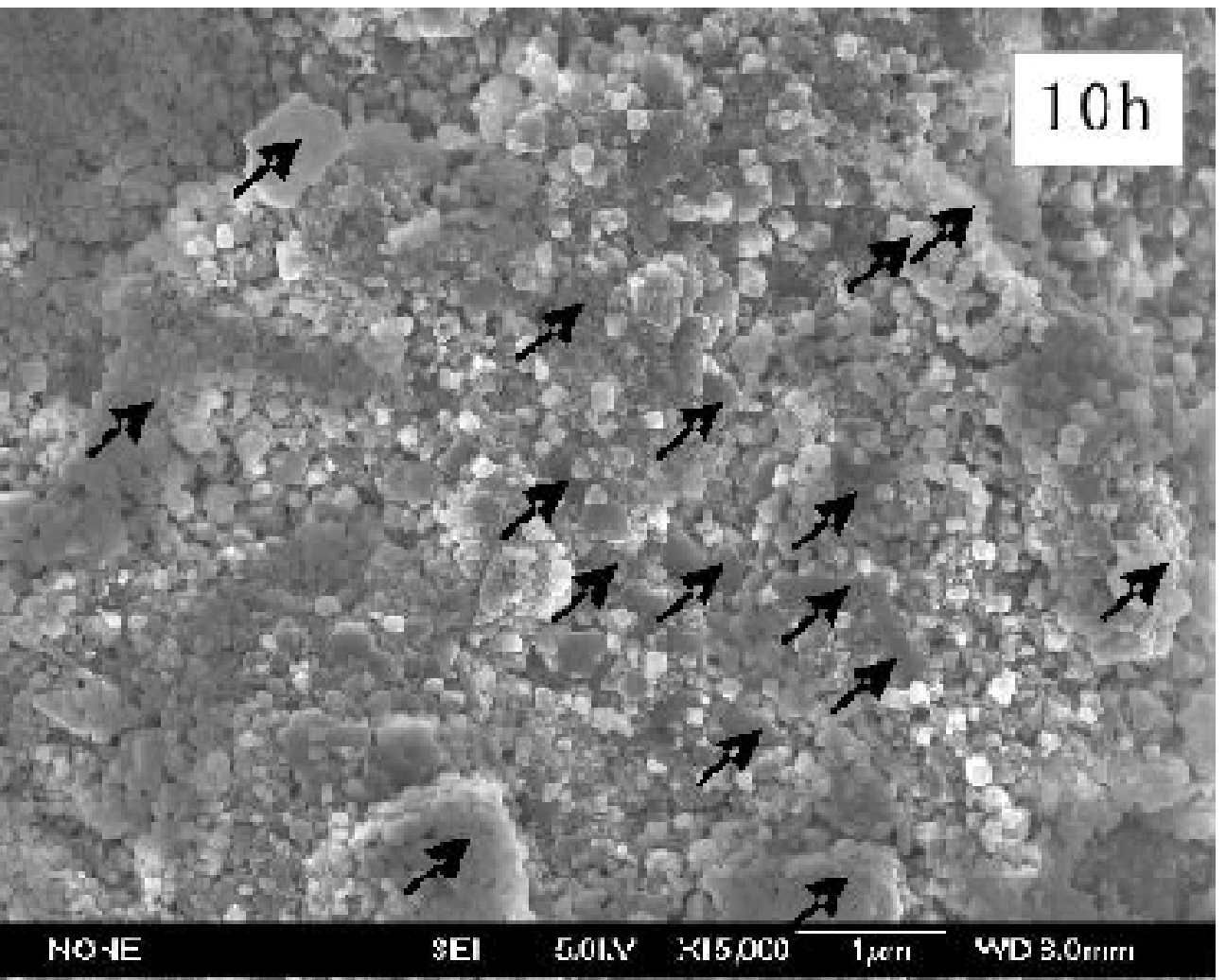}
\end{figure}

\newpage
Fig.2
\begin{figure}[hp]
\centering
\includegraphics[width=180pt]{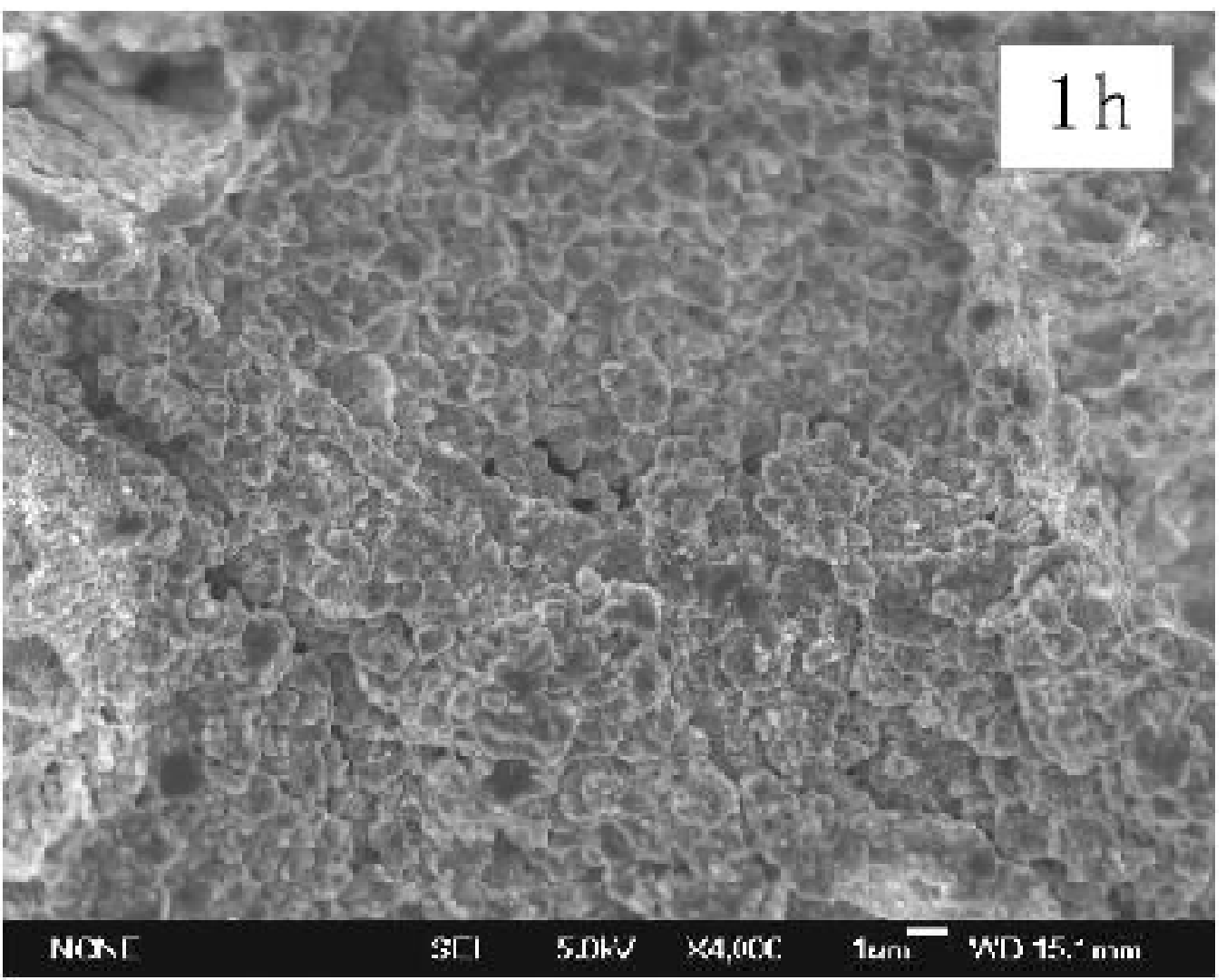}%
\includegraphics[width=180pt]{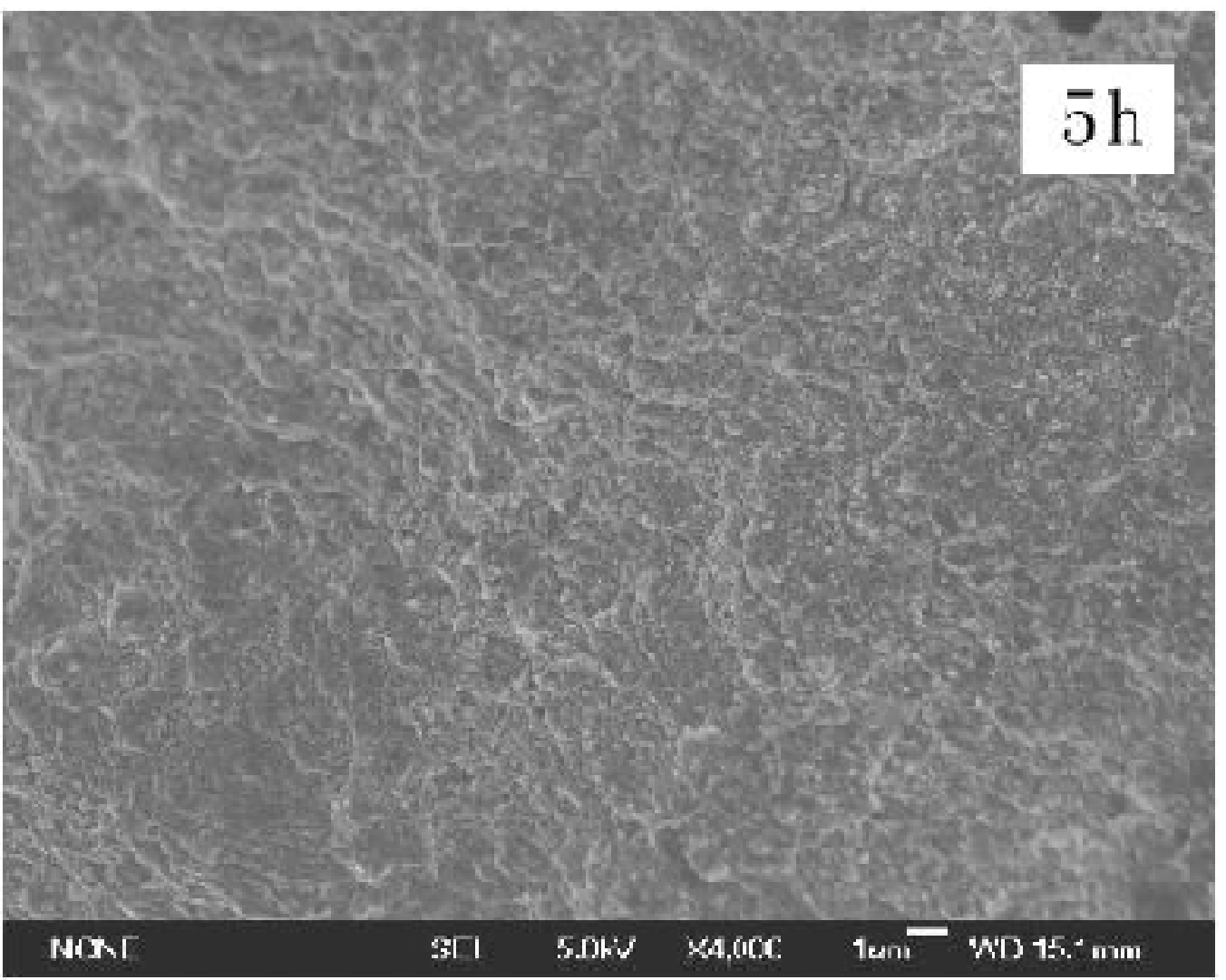}
\includegraphics[width=180pt]{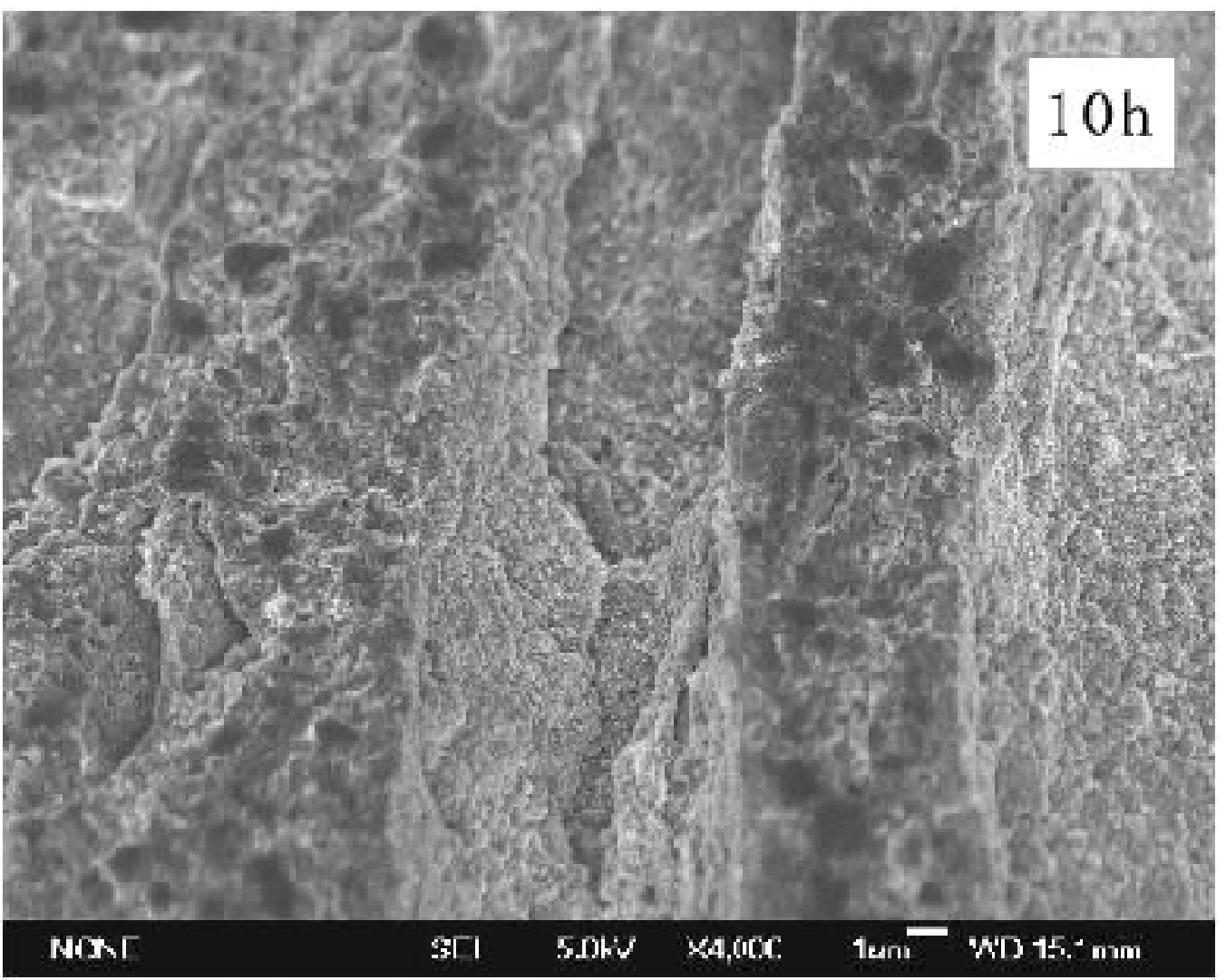}
\end{figure}

\newpage
Fig.3
\begin{figure}[hp]
\centering
\includegraphics[width=300pt]{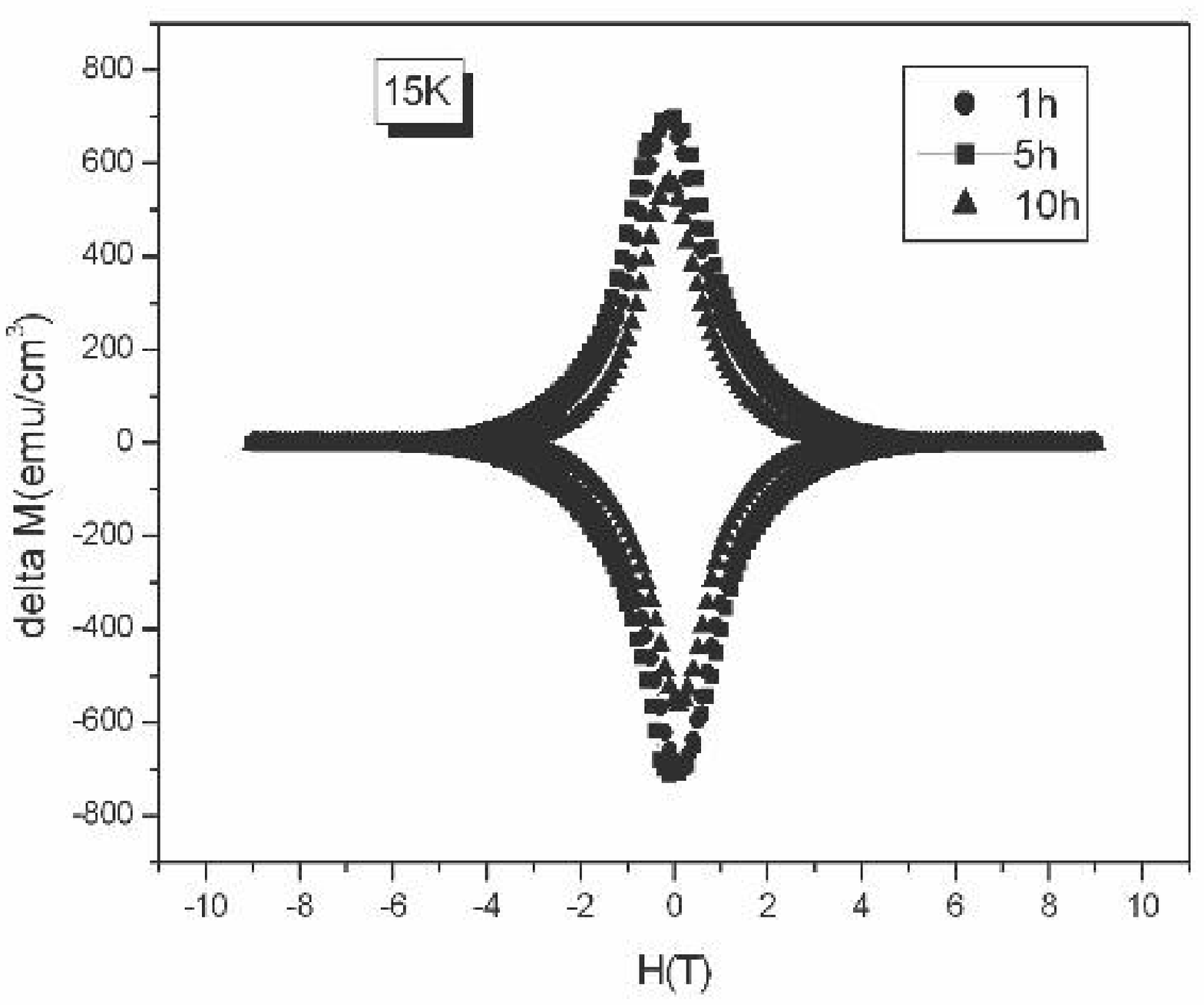}
\end{figure}

\newpage
Fig.4
\begin{figure}[hp]
\centering
\includegraphics[width=300pt]{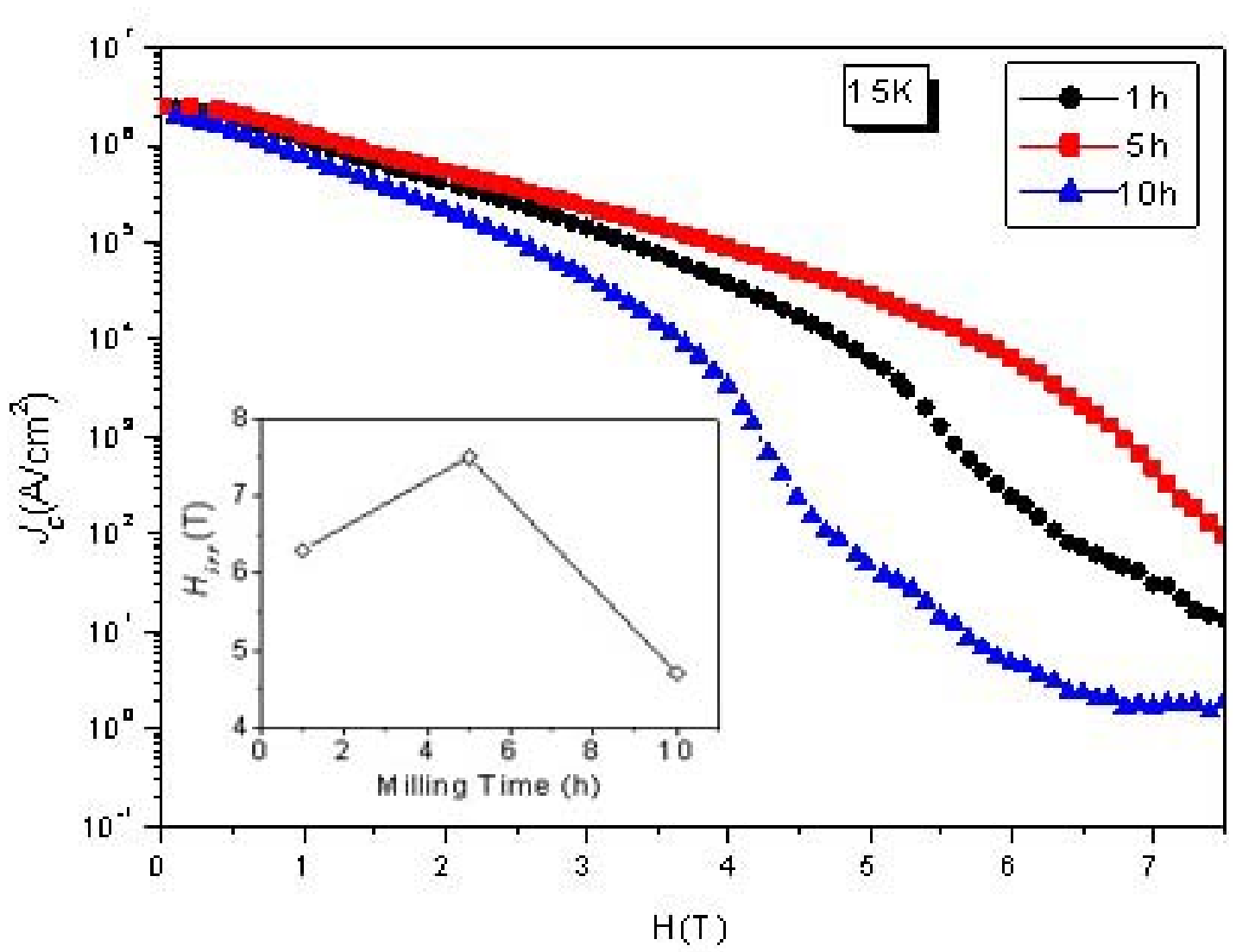}
\end{figure}

\newpage
Fig.5
\begin{figure}[hp]
\centering
\includegraphics[width=300pt]{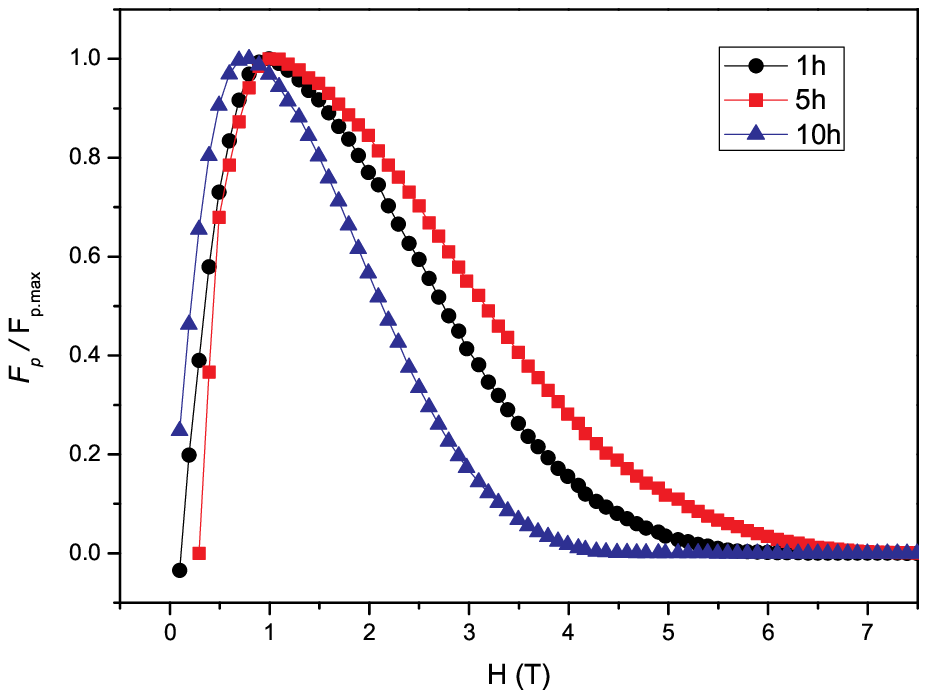}
\end{figure}

\newpage
Fig.6
\begin{figure}[hp]
\centering
\includegraphics[width=400pt]{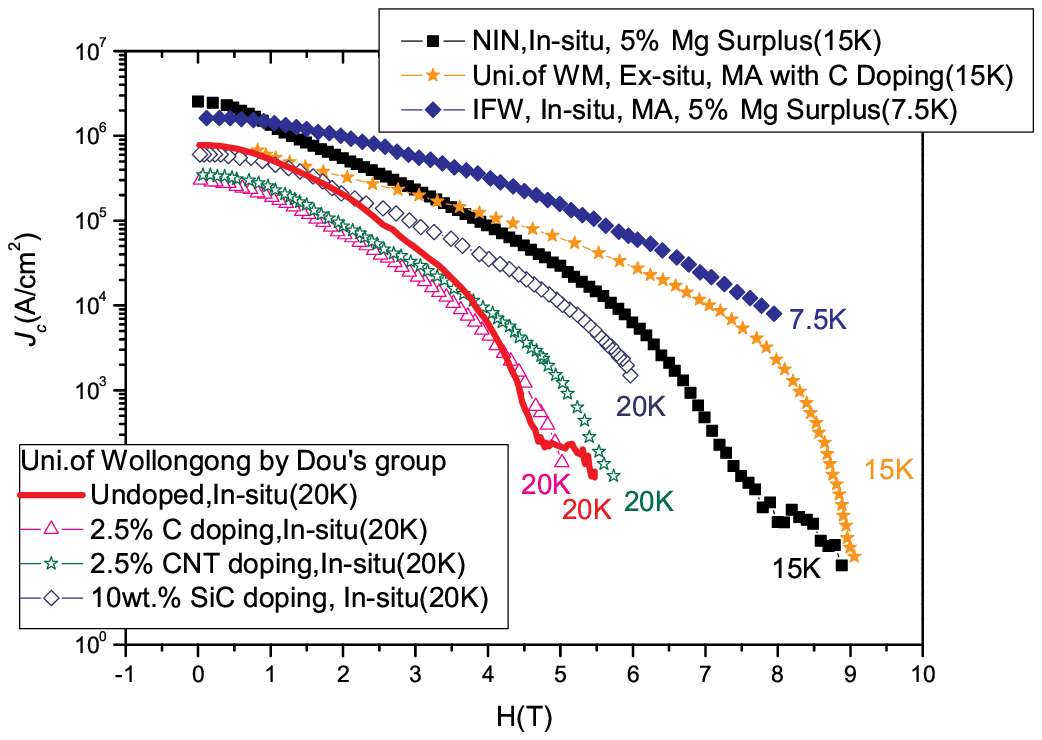}
\end{figure}


\section*{References}
\begin{thebibliography}{25}
\bibitem{book1} D.C.Larbalestier, L. D. Cooley, M. O. Rikel, A. A. Polyanskii,
J. Jiang, S. Patnaik, X. Y. Cai, D.M. Feldman, A. Gurevich, A.A. Squitieri,
M.T. Naus, C. B. Eom, E.E. Hellstrom, R. J. Cava, K.A. Regan, N. Rogadao,
M. A. Hayward, T. He, J. S. Slusky, P. Khalifah,  K. Inumaru, and M. Haas,
Nature 410(2001), pp.186-189.

\bibitem{book2} Y. Bugoslavsky, G. K. Perkins, X. Qi, L. F. Cohen, and A. D.
Caplin, Nature, 410(2001), pp.563-565.

\bibitem{book3} Y.Bugoslavsky, L.F.Cohen, G.K. PERKINS, M. Polichetti, T.J. Tate, R.
Gwilliam, and A. D. Caplin, Nature, 411(2001), pp.561-563.

\bibitem{book4} V.N.Narozhnyi, G. Fuchs, A, Handstein, A. Gumbel, J. Eckert, K.
Nenkov, D. Hinz,    O. Gutfleisch, A. Walte, L. N. Bogacheva, I. E.
Kostyleva, K.-H. Muller, L. Schultz,   Int. Conf. on
Superconductivity, CMR \& Related Materials: Novel Trends
(SCRM2002), Giens, France, 1-8 June 2002

\bibitem{book5} G. Fuchs, K.-H. Muller, A. Handstein, K. Nenkov, V.N. Narozhnyi, D. Echert,
M. Wolf and L. Schultz, Solid State Commun., Vol. 118, pp.497-501, 2001.

\bibitem{book6} A. Gumbel, O. Perner, J. Eckert, G. Fuchs, K. Nenkov, K.-H. Muller,
and L. Schultz, IEEE TRANSACTIONS ON APPLIED SUPERCONDUCTIVITY, Vol.
13, No. 2, JUNE 2003.

\bibitem{book7} A. Gumbel, J. Eckert, G. Fuchs, K. Nenkov, K.-H. Muller, and L. Schultz,
APPLIED PHYSICS LETTERS, Vol. 80, No. 15, 15 APRIL 2002, pp.2725-2727.

\bibitem{book8} W. Ha¦Âler, C. Roding, C. Fischer, B. Holzapfel, O. Perner, J. Eckert,
K. Nenkov and G. Fuchs, Supercond. Sci. Technol. 16 (2003)
pp.281-284.

\bibitem{book9} C. Fischer, C. Roding, W. Ha¦Âler, O. Perner, J. Eckert, K. Nenkov, G. Fuchs,
H. Wendrock, B. Holzapfel and L. Schultz, APPLIED PHYSICS LETTERS,
Vol. 83,   No. 9, 1 SEPTEMBER 2003, pp.1803-1805.

\bibitem{book10} O. Perner, J. Eckert, W. Ha¦Âler, C. Fischer, K-H Muller, G. Fuchs,
 B Holzapfel and L Schultz, Supercond. Sci. Technol. 17 (2004) pp.1148-1153.

\bibitem{book11} Olaf Perner, Wolfgang Ha¦Âler, Claus Fischer, Gunter Fuchs,
Bernhard Holzapfel, Ludwig Schultz and Jurgen Eckert,
IEEE TRANSACTIONS ON APPLIED SUPERCONDUCTIVITY, VOL. 15, NO. 2, JUNE 2005.

\bibitem{book12} O. Perner, J. Eckert, W. Ha¦Âler, C. Fischer, J. Acker, T. Gemming, G. Fuchs,
B. Holzapfel and L. Schultz, JOURNAL OF APPLIED PHYSICS 97,
056105(2005), pp.1-3.

\bibitem{book13} T. Massalski, Ed., Binary Alloy Phase Diagrams,
2nd ed. Materials Park, OH: ASM International, 1990.

\bibitem{book14} A. Serquis, L. Civale, J. Y. Coulter, D. L. Hammon, X. Z. Liao,
Y. T. Zhu, D. E. Peterson, F. M. Mueller, V. F. Nesterenko, and S.
S. Indrakanti, Preprint cond-mat/0404052.

\bibitem{book15} Sihai Zhou, Alexey V. Pan, Huakun Liu, Shixue Dou,
Physica C 382 (2002), pp.349-354.

\bibitem{book16} H. Fang, P. Gijavanekar, Y. X. Zhou, G. Liang, P. T. Putman, and K. Salama,
IEEE TRANSACTIONS ON APPLIED SUPERCONDUCTIVITY, VOL. 15, NO. 2, JUNE 2005.

\bibitem{book17}  Y. F. Wu, Y. Feng, G. Yan, J. S. Li, H. P. Tang, S. K. Chen, Y. Zhao, M. H. Pu,
H. L. Xu, C. S. Li,Y. F. Lu, Preprint cond-mat/0603536.


\bibitem{book18}Y. Zhao, C.H. Cheng, Y. Feng, T. Machi, D.X. Huang, L. Zhou, N. Koshizuka,
M. Murakami, Physica C 386 (2003), pp.581-587.

\bibitem{book19} R. Flukiger , H.L. Suo, N. Musolino, C. Beneduce, P. Toulemonde,
P. Lezza, Physica C 385 (2003) pp.286-305.

\bibitem{book20} H Fujii, K Togano and H Kumakura,
Supercond. Sci. Technol. 16 (2003), pp.432-436.

\bibitem{book21} B. J. Senkowicz, J. E. Giencke, S. Patnaik, C. B. Eom, E. E. Hellstrom,
and D. C. Larbalestier, Preprint cond-mat/0411199.

\bibitem{book22} O. Perner, J. Eckert, W. H??ler, C. Fischer, J. Acker, T. Gemming, G. Fuchs,
B. Holzapfel, and L. Schultz, JOURNAL OF APPLIED PHYSICS 97, 056105,
2005.

\bibitem{book23}S. X. Dou, S. Soltanian, J. Horvat, X. L. Wang, S. H. Zhou,
M. Ionescu, and H. K. Liu£¬APPLIED PHYSICS LETTERS£¬VOLUME 81,
NUMBER 18£¬28 OCTOBER 2002, pp.3419-3421.

\bibitem{book24} S. Soltanian, J. Horvat, X.L. Wang, P. Munroe,
S.X. Dou, Physica C 390 (2003) pp.185-190.

\bibitem{book25} S. X. Dou, W. K. Yeoh, J. Horvat, and M. Ionescu,
APPLIED PHYSICS LETTERS, VOLUME 83, NUMBER 24, 15 DECEMBER 2003.

\end{thebibliography}
\end{document}